\documentclass[prl,twocolumn,aps,superscriptaddress,showpacs,floatfix]{revtex4}
\usepackage{amsmath}
\usepackage{graphicx}

\begin{document}
\title{Universal and reconfigurable logic gates in a noise-triggered artificial neuron node}
\author{L. Worschech} 
\author{F.  Hartmann} 
\author{A. Forchel} 
\affiliation{Technische Physik, Physikalisches Institut, Universit\"at W\"urzburg, Am Hubland, 97074 W\"urzburg, Germany}
\author{J. Ahopelto} 
\affiliation{VTT Micro and Nanoelectronics, P.O. Box 1000, FI-02044 VTT Espoo, Finland}
\author{I. Neri}
\author{L. Gammaitoni}
\affiliation{NiPS Laboratory, Dipartimento di Fisica, Universit\`a di Perugia,
I-06123 Perugia, Italy, and Istituto Nazionale di Fisica Nucleare,
Sezione di Perugia, I-06123 Perugia, Italy}
\date{\today}

\begin{abstract}
Submicron-sized mesas of resonant tunneling diodes (RTDs) with split drain contacts have been realized and the current-voltage characteristics have been studied in the bistable regime at room temperature. Dynamically-biased, the RTDs show noise-triggered firing of spike-like signals and can act as reconfigurable universal logic gates for small voltage changes of a few mV at the input branches. These observations are interpreted in terms of a stochastic nonlinear processes in the split RTD, which are found to be robust against noise. The split RTDs show also gain for the fired-signal bursts, can be easily integrated to arrays of multiple inputs and have thus the potential to mimic neuron nodes in nanoelectronic circuits.
\end{abstract}
 \maketitle
Neurons handle and fire spike-like signal trains in a noisy environment. How they can code, encode and compute information massively affected by noise is still one of the most fascinating and unresolved questions in information science. Indeed, the success of computing digital signals in microelectronics is exactly based on the fact to avoid noise by boosting signals to digital levels well separated from noise margins. However, this noise paradigm has to be reconsidered in future signal processing as signal amplitudes and noise floor will approach each other \cite{Hanson,[5],[6],luca-apl,quantum1, quantum2, [3], Sano, Kish}.

\begin{figure}[ht]
\includegraphics*[width=9.0cm]{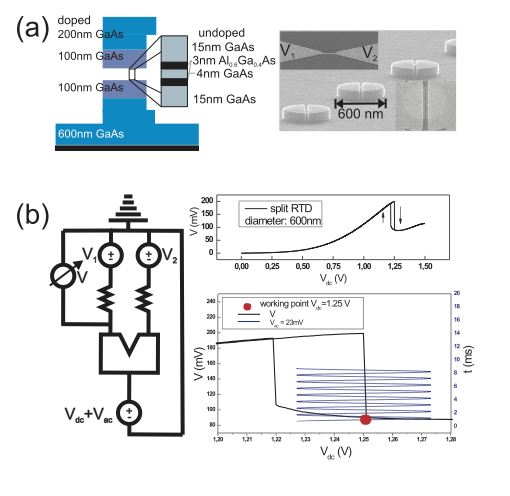}
\caption{(a) Left part: Sketch of layer sequence. Molecular beam epitaxy was applied to grow 3 nm thick AlGaAs barriers embedding a 4 nm thick GaAs layer. The outer GaAs layers were doped with Si to form contacts, whereas the inner parts are undoped. Right part: Electron beam lithography and etching were applied to define split RTD mesas with a diameter of 600 nm. The top branches were contacted by gold contacts separately. Different voltages were applied to the top RTD branches ($V_1$ and $V_2$). (b) Left part: Measurement circuit diagram. Voltages $V_1$ and $V_2$ were applied in series with 2k$\Omega$ resistors to the RTD branches. The working point voltage $V_{dc}$ + $V_{as}$ was applied to the back contact. The voltage $V$ was measured as indicated. Right part: The I-V trace of the RTD shows a bistable transition with a hysteresis of about 30 meV. The working point was chosen in such a way that $V_{dc}=1.25 V$ and $V_{ac}$ was modulated between 20 and 30 mV with a frequency of 1 kHZ. $V_1$ and $V_2$ were set to 0 or 4 mV.}
\label{fig:1} 
\end{figure}

Different routes are possible to compute noise affected signals, e.g. probabilistic strategies \cite{ [9]} and nonlinear stochastic dynamics \cite{Ditto} in order to design new kinds of logic gates. Interestingly, by applying principles associated with the well known Stochastic Resonance phenomenon, it is even possible to take advantage by the presence of noise \cite{SR}. Enabling requirements are bistable switching with threshold voltages so small that noise can activate the switching. Here we present an approach based on dynamically modulated bistable switching in resonant tunneling diodes (RTD) with branched drains as nanoelectronic devices, which generate spike-like signal trains efficiently controlled by the electronic environment. A novel noise-triggered operating scheme as a universal and reconfigurable logic gate is demonstrated and related to a nonlinear stochastic process. Such a scheme is shown to operate properly in the presence of a significant amount of noise.

In the upper part of Fig.\ref{fig:1} a sketch of the RTD layer sequence is shown. The structure was grown by molecular beam epitaxy. The central part of the RTD is built up by two 3 nm thick AlGaAs barriers embedding a 4 nm thick GaAs layer. Afterward, electron beam lithography and etching were applied for the definition of small mesas with a trench through the upper doped layers. Gold contacts were realized to independently contact the two upper RTD branches further used as input terminals (right part of Fig.1(a)). The measurement circuit diagram is depicted in Fig.\ref{fig:1}(b). The samples were tested at room temperature in the dark. The RTD shows a resonant peak in the I-V curve at $1.25$ $V$. A hysteresis of $30$ $mV$ was found between the up and down sweep of $V_{dc}$.

The RTD nanojunctions were tested in the following way (see Fig. \ref{fig:1}(b)). At the back contact a working point voltage $V_{dc}=1.25$ $V$ was applied superimposed by an ac signal $V_{ac}$ with a frequency of $1$ $kHz$. The left and right upper branches were used as inputs with voltages $V_1$ and $V_2$, respectively. In the right part of Fig.1(b) the working principle is presented. Via the voltage $V_{dc}$ the RTD was driven close to the bistable transition. When the amplitude $V_{ac}$ is ramped up close to a critical threshold voltage, noise-triggered switching can drive the RTD in the upper state. The system is then driven back again to the lower state due to the time-dependent modulation. In the following, it is demonstrated that small voltage changes at the input branches of only a few mV lead to pronounced switching of spike-like signal trains.

\begin{figure}[ht]
\includegraphics*[width=9.0cm]{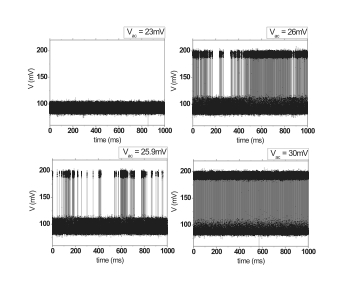}
\caption{Time series of the RTD output voltage $V$ for different ac voltages $V_{ac}$. For $V_{ac}=23$ $mV$, the RTD stays in the lower state. By increasing $V_{ac}$ to $25.9$ $mV$ it eventually happens that the RTD switches to the upper state. Then a shift of only $0.1$ $mV$ increases efficiently the number of spike-like switching events. For $V_{ac}=30$ $mV$. The RTD stays equally in the upper and the lower state. The mean is used  here as a measure for the output signal.}
\label{fig:2} 
\end{figure}

In Fig.\ref{fig:2} time traces of the RTD output voltage $V$ are shown. The bistable character of the dynamics is apparent. By passing from $V_{ac}=23$ to $30$ $mV$ the system is changed totally by keeping constant all the other parameters. For $V_{ac}=23$ $mV$ the RTD stays in the lower state with $V<105$ $mV$. For $V_{ac}=25.9$ $mV$ the RTD is close to the threshold so that noise-induced random jumps from one state to the other appear. By increasing $V_{ac}$ now by only $0.1$ $mV$, bursts of spike-like signal trains can be observed. For $V_{ac}=30$ $mV$ the upper and the lower state of the RTD are almost equally occupied.

\begin{figure}[ht]
\includegraphics*[width=9.0cm]{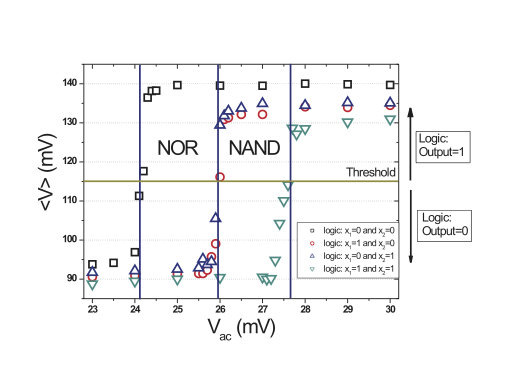}
\caption{$\langle V \rangle$ interpreted as a logic output $0$ for values smaller than $115$ $mV$ or $1$ for values  exceeding $115$ $mV$ as a function of the inputs: $V_{i}$ = $0$ ($0$ $mV$), $V_{i}$ = 1 ($4$ $mV$) for different voltages $V_{ac}$.}
\label{fig:3} 
\end{figure}

In Fig.\ref{fig:3} we show the mean value of $V$ as a function of $V_{ac}$ for different input parameter voltages $V_1$ and $V_2$ associated as logic $0$ and $1$ for $V_{1,2} = 0$ $mV$ and $4$ $mV$, respectively. A threshold of 115 mV is defined above which the mean $V$ is interpreted as a logic output $1$ else $0$. We stress the fact that a fairly large output change is produced as a function of a relatively small change in the input. Based on these results we can interpret the functioning of the split RTD in terms of a universal NAND logic gate. Most interestingly the logic functioning of such a gate can be easily changed into that of a logic NOR gate. In fact, as we show in Fig.\ref{fig:3} by changing the amplitude of the periodic forcing from $V_{ac}=25.0\pm0.5$ mV to $V_{ac}=26.5\pm0.5$ the logic behavior changes from NOR to NAND. Such a change happens in a relatively small amplitude range (within $10\%$) of the periodic forcing.

In the following we propose a stochastic nonlinear process to model the observed logic function of the split RTD functioning. The model is also capable of predicting a very useful property, i.e. a wide tolerance to the noise affecting the input signals.
We start with considering a stochastic dynamic equation, Langevin kind, for the output quantity, that we will call here $v(t)$, subjected to a bistable static potential $U(v)$ and a time dependent force $F(t)$. 

The time dependent force is composed by three different components: 

\begin{equation}
F(t) = A(t)+\xi(t)+I(t)
\end{equation}

where $A(t) = A~sin(\omega t)$ is a time periodic signal that, without loss of generality, can be assumed harmonic. $\xi(t)$ is a 
stochastic force that mimics the presence of noise, that can be assumed exponentially correlated, Gaussian distributed, with zero mean and standard deviation $\sigma$.  $I(t) = v_1(t) + v_2(t)$ represents the input signal composed by the composition of the two inputs at the two split RTD branches. $v(t)$ represents here the split RTD output whose dynamics can be described as:

\begin{equation}
\dot{v} =  \frac{dU(v)}{dv} + A(t) +\xi(t) + d\ I(t)
\label{lang1}
\end{equation}

To fix our ideas let's consider the quartic {\em double well} potential, but what we are going to say is generically valid for any bistable potential, regardless of the specific analytic form:

\begin{equation}
U(v) =  -a \frac{1}{2}v^2 + b \frac{1}{4}v^4 + cv 
\label{P1a}
\end{equation}
where $a$,$b$, $c$ and $d$ are suitable constant coefficients. 

When $c=0$ the potential is symmetric and in the absence of any external force, i.e. $F(t) = 0$, the system output $v(t)$ can assume with equal probability one of the two values $\pm v_m$ corresponding to the two stable minima of the potential $U(v)$. When $c \neq 0$, the potential is not symmetric anymore and it appears tilted in one direction or the other depending on the sign of $c$. If $|c| \geq c_{th}$ the potential becomes monostable.  When $F(t) \neq 0$ the system output perform a rich dynamics composed mainly by small oscillations around one of the two minima and occasional jumps from one minima to the other, depending on the values of the instant force and of the potential parameters.

\begin{figure}[ht]
\includegraphics*[width=9.0cm]{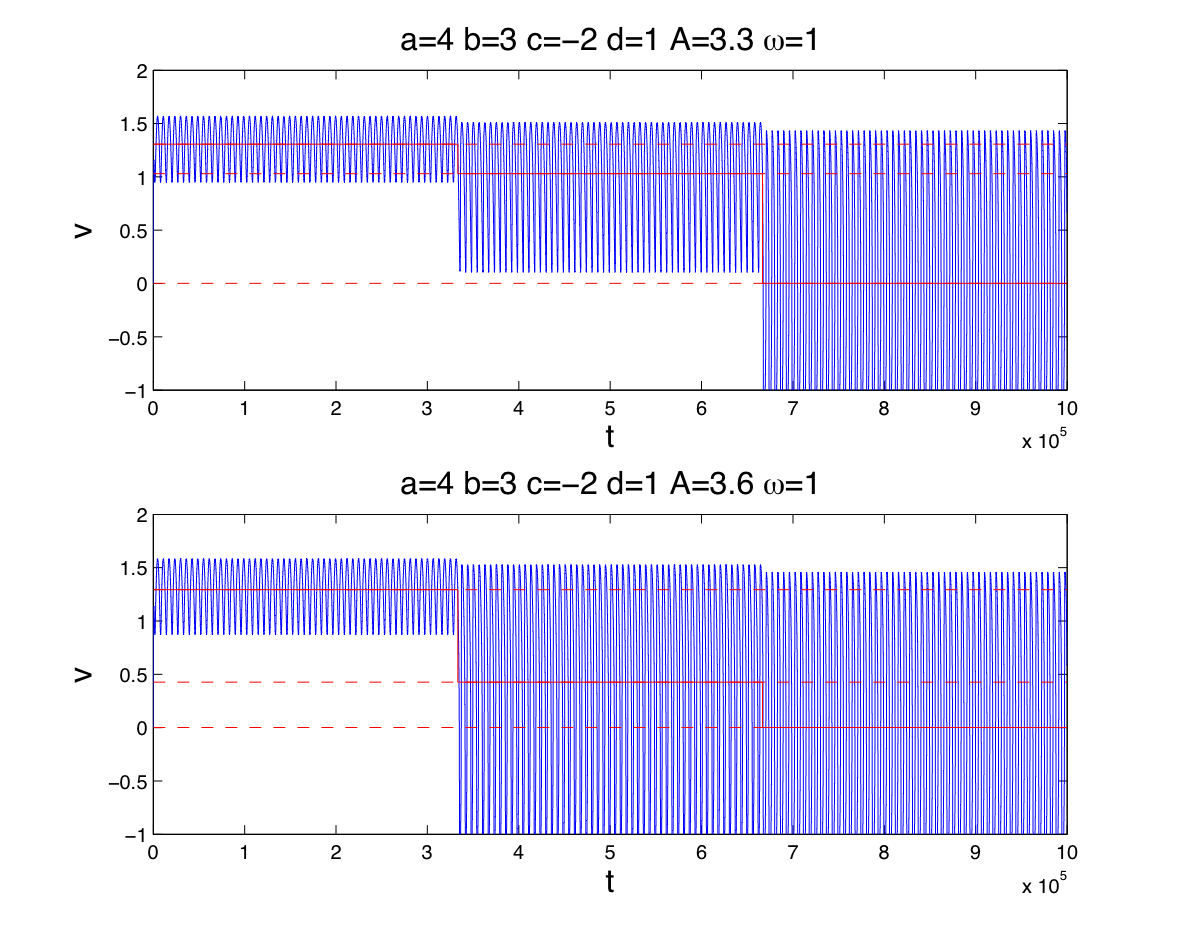}
\caption{
$v(t)$ for three different combinations of the input values $I(t) = v_1(t) + v_2(t) = 0,1,2$. Here $\sigma = 0$. Upper panel $A=3.3$, if, as an example, logical $1$ is assumed for $\langle v \rangle \geq 0.5$ and the logical $0$ for  $\langle v \rangle < 0.5$, the NAND behavior is reproduced. Lower panel $A=3.6$, NOR behavior is reproduced.}
\label{fig:4} 
\end{figure}

\begin{figure}[ht]
\includegraphics*[width=9.0cm]{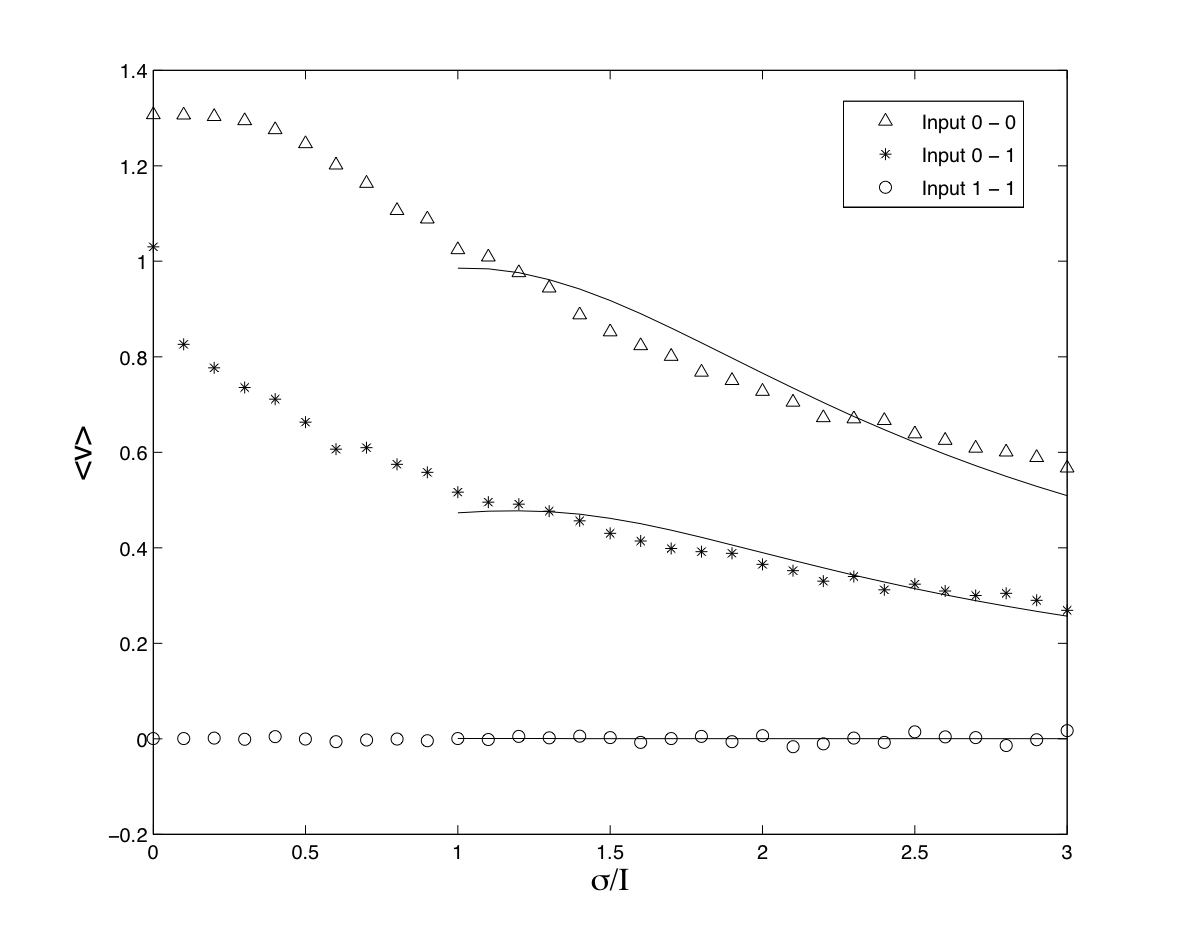}
\caption{
NAND configuration, $A=3.3$. $\langle v \rangle$ for $I(t) = 0,1,2$, as a function of the noise standard deviation $\sigma/I$ where $I$ is the amplitude of the single signal input (here $I=1$). The continuous lines represent the theoretical prediction.
}
\label{fig:5} 
\end{figure}

Initially we focus on the behavior illustrated in Fig.\ref{fig:2} where the bistable character of the RTD dynamics is exploited in order to perform the NAND/NOR gate functions. In Fig.\ref{fig:4} we show the time series $v(t)$ for three different combinations of the input values $I(t) = v_1(t) + v_2(t) = 0,1,2$. The output $v(t)$ shows an oscillating character, qualitatively similar to $V$ in the right parts of Fig.\ref{fig:2}. Notably, in the upper panel, $A=3.3$ when the input changes from $I(t) = v_1(t) + v_2(t) = 0 + 0 = 0$ to $I(t) = 1 + 0 = 0 + 1 = 1$ to $I(t) = 1 + 1 = 2$ the mean output $\langle v \rangle$ switches from $\langle v \rangle = 1.3$, to $\langle v \rangle = 1.0$, to $\langle v \rangle = 0.0$. Thus, if we associate the logic state $1$ to $\langle v \rangle \geq 0.5$ and the logic state $0$ to  $\langle v \rangle \leq 0.5$ we have easily reproduced the NAND behavior as shown by the RTD in Fig.\ref{fig:3}. In the lower panel we show the output $v(t)$ for the same conditions of the upper panel but with a value of $A=3.6$. In this case the mean output $\langle v \rangle$ switches from $\langle v \rangle = 1.3$, to $\langle v \rangle = 0.4$, to $\langle v \rangle = 0.0$ thus reproducing the NOR behavior. At this point it is easy to compare these results to those in Fig.\ref{fig:4} where the change in logic function is obtained for a $10\%$ change of the value of the amplitude of the periodic forcing.

The present strategy aimed at reducing the dissipated power in traditional logic gates based on transistors requires a decrease in the transistors operative voltage thus increasing the vulnerability to the noise. In the following we show that the new scheme just presented for the description of the RTD functioning, shows a significant tolerance to the noise. In Fig.\ref{fig:5} we show $\langle v \rangle$ as a function of the noise standard deviation $\sigma$ for the NAND configuration. The continuous line refers to a theoretical prediction that can be easily obtained from:

\begin{equation}
\langle v \rangle = \int_0^T \frac{N(t)}{T}~ dt \int_{-\infty}^{+\infty} v~ e^{- \frac{U(v)+v (A(t)+I(t))}{\sigma^2 \tau}} dv
\end{equation}
where 

\begin{equation}
N(t) \propto \frac{1}{\int_{-\infty}^{+\infty} e^{- \frac{U(v)+v (A(t)+I(t))}{\sigma^2 \tau} }dv}
\end{equation}

As it can be seen in the figure a logical output  coherent with the NAND scheme (logical $1$ corresponding to $\langle v \rangle \geq 0.5$) is maintained also for noise intensities comparable with the value of the single input. This is quite a remarkable property that could allow the operation of the logic gate in unusually high noise environments. Moreover, with reference to Fig.\ref{fig:5} we notice that a lower choice of the threshold value for the logical $0$ might allow the operation of the gate also in the presence of a much larger noise. However in this case the re-configurability of the gate from NAND to NOR might be compromised \cite{future}.

\begin{figure}[ht]
\includegraphics*[width=9.0cm]{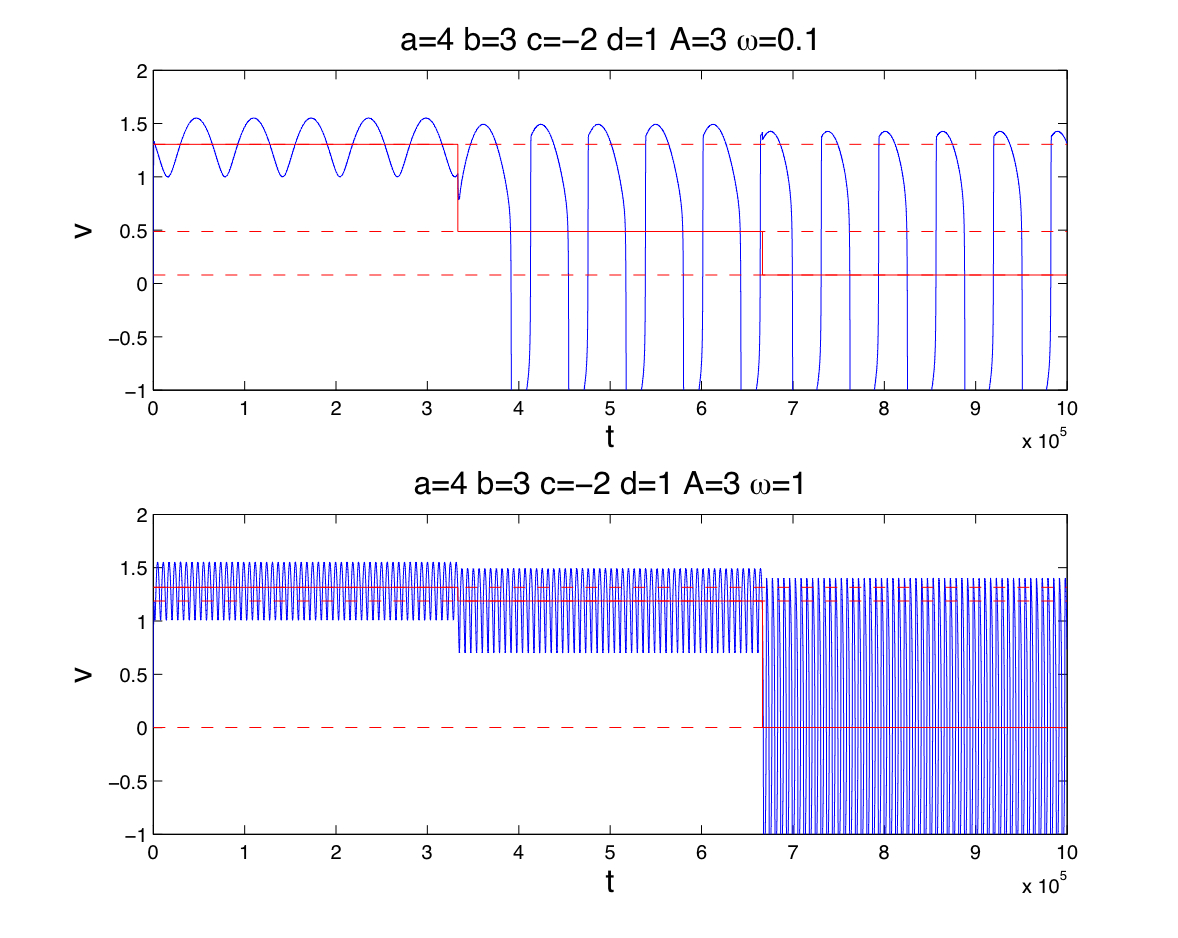}
\caption{
$v(t)$ for three different combinations of the input values $I(t) = I1(t) + I2(t) = 0,1,2$. Here $\sigma = 0$, $A=3.3$. Upper panel $\omega = 0.1$. As previously observed due to the value assumed for $\langle v \rangle$ we observe a NAND behavior. Lower panel $\omega = 1.0$, NOR behavior is reproduced.}
\label{fig:6} 
\end{figure}

Finally we note that in order to change the configuration from NAND to NOR it is also possible to act on parameters other than $A$. One example is the frequency $\omega$. In Fig.\ref{fig:6} we show the time series $v(t)$ for three different combinations of the input values as in Fig.\ref{fig:4}. Another possibility to reconfigure the gate behavior is to act on the control parameter $c$ in eq.(3). Also in this case a small change on this parameter allow a switch between the two logics\cite{future}.

In conclusion we have presented a novel concept device based on the Resonant Tunneling effect that can be usefully employed as a reconfigurable universal logic gate. The gate can be set to behave as a NAND gate or a NOR gate by simply changing the amplitude of the periodic signal ($10\%$), its frequency or a DC bias voltage. We have proposed an analytical model based on stochastic nonlinear dynamics that is capable of reproducing all the features experimentally observed on the RTD. Moreover, digital simulations operated on the analytical model indicates that the device functioning is tolerant to noise as large as $100 \%$ of the digital input.

The authors gratefully acknowledge financial support from European Commission (FPVI, STREP Contract N. 034236 SUBTLE: Sub KT Low Energy Transistors and Sensors).

\end{document}